\documentclass[12pt]{article}
\begin{document}
\setlength{\baselineskip}{0.30in}
\newcommand{\nc}{\newcommand}
\newcommand{\beq}{\begin{equation}}
\newcommand{\eeq}{\end{equation}}
\newcommand{\be}{\begin{eqnarray}}
\newcommand{\ee}{\end{eqnarray}}
\newcommand{\num}{\nu_\mu}
\newcommand{\nue}{\nu_e}
\newcommand{\nut}{\nu_\tau}
\newcommand{\nus}{\nu_s}
\newcommand{\mnus}{m_{\nu_s}}
\newcommand{\taus}{\tau_{\nu_s}}
\newcommand{\nnt}{n_{\nu_\tau}}
\newcommand{\rnt}{\rho_{\nu_\tau}}
\newcommand{\mnt}{m_{\nu_\tau}}
\newcommand{\tnt}{\tau_{\nu_\tau}}
\newcommand{\bi}{\bibitem}
\newcommand{\rar}{\rightarrow}
\newcommand{\lar}{\leftarrow}
\newcommand{\lrar}{\leftrightarrow}
\newcommand{\dm}{\delta m^2}
\newcommand{\so}{\, \mbox{sin}\Omega}
\newcommand{\co}{\, \mbox{cos}\Omega}
\newcommand{\sotil}{\, \mbox{sin}\tilde\Omega}
\newcommand{\cotil}{\, \mbox{cos}\tilde\Omega}
\newcommand{\raa}{\rho_{aa}}
\newcommand{\rss}{\rho_{ss}}
\newcommand{\rsa}{\rho_{sa}}
\newcommand{\ras}{\rho_{as}}
\makeatletter
\def\alt{\mathrel{\mathpalette\vereq<}}
\def\vereq#1#2{\lower3pt\vbox{\baselineskip1.5pt \lineskip1.5pt
\ialign{$\m@th#1\hfill##\hfil$\crcr#2\crcr\sim\crcr}}}
\def\agt{\mathrel{\mathpalette\vereq>}}

\newcommand{\eq}{{\rm eq}}
\newcommand{\tot}{{\rm tot}}
\newcommand{\M}{{\rm M}}
\newcommand{\coll}{{\rm coll}}
\newcommand{\ann}{{\rm ann}}
\makeatother

\begin{center}
\vglue .06in
{\Large \bf { Restrictions on neutrino oscillations from BBN.
Non-resonant case.
}}
\bigskip
\\{\bf A.D. Dolgov
\footnote{Also: ITEP, Bol. Cheremushkinskaya 25, Moscow 113259, Russia.}
 \\[.05in]
{\it{INFN section of Ferrara\\
Via del Paradiso 12,
44100 Ferrara, Italy}
}}
\end{center}

\begin{abstract}

New nucleosynthesis bounds on the oscillation parameters of active 
neutrinos mixed with a sterile one are derived for the non-resonant 
case. The controversy in the literature whether to use the 
annihilation rate or the total reaction rate for the estimates 
of sterile neutrino production is resolved in favor of the
annihilation rate. In contrast to previous papers, the restrictions
on oscillations of electronic neutrinos are weaker than those of
muonic and tauonic ones.

\end{abstract}

Neutrino oscillations in the early universe, especially if active
neutrinos
are mixed with sterile ones, would have a noticeable impact on primordial 
nucleosynthesis~\cite{dolgov81} and this permits to obtain interesting
restrictions on the oscillation parameters. The results very much depend
upon a possible MSW resonance transition~\cite{mikheev85} in the primeval 
plasma. In the case of resonance ($\dm <0$) the oscillations are much
more efficient, neutrino spectrum can be strongly distorted, and the
lepton asymmetry in the sector of active neutrinos can be enhanced by
several orders of magnitude. However the calculations in this case 
are very complicated and controversial conclusions have been reached
by different groups. For the discussion and the list of references
see the recent papers~\cite{dolgov99,dibari99}. In the non-resonant case
the calculations are much simpler but there is also a disagreement
between the papers~\cite{barbieri90} and all the subsequent 
works, see e.g.~\cite{kainulainen90}-\cite{shi93}. In the first papers
it was assumed that the probability of production of sterile neutrinos,
$\nus$, is proportional to the rate of (inverse) annihilation of
active neutrinos into light fermions in the plasma, 
$\gamma_{ann}$, while in all other papers it was argued that the
rate of production of $\nus$ is proportional to a much larger total
scattering rate, $\gamma_{tot} = \gamma_{el}+\gamma_{ann}$. 
Correspondingly the BBN (big bang nucleosynthesis) bounds on the
oscillation parameters would be much more restrictive. 

It is shown here that these arguments are not correct and the 
probability of $\nus$ production is indeed proportional to the
annihilation rate in agreement with ref.~\cite{barbieri90}, though
more precise calculations presented below result in stronger bounds
than found in the papers~\cite{barbieri90} but still weaker than those
obtained in the above quoted papers~\cite{kainulainen90}-\cite{shi93},
where the probability of production  
was taken to be proportional to $\gamma_{tot}$. The simple argument
showing that the effect of oscillations vanishes in the limit of
$\gamma_{ann}=0$ is that in this limit the total number density
of active and sterile neutrinos is conserved, $n_a + n_s = const$,
(see below eqs.~(\ref{draa},\ref{drss})) and roughly speaking their total 
energy density remains the same as in the absence of oscillations. 
In fact the situation is somewhat more complicated because at an 
early hot stage the equilibrium with respect  to annihilation was 
also reached, so the obtained expressions do not permit to take the 
limit of vanishing $\gamma_{ann}$.

The difference between the approach of the present paper and 
all the other ones, where the BBN bounds have
been derived, is that in those papers the breaking of coherence
of the oscillations was described by the simplified anzats
to the r.h.s. of the kinetic equations::
\be
-\gamma \, \{g^2,\rho-\rho^{(eq)} \}  
\label{grho}
\ee
where curly brackets denote anti-commutator and
$g$ is the interaction matrix. In flavor basis it has the only
nonzero entry in (a,a)-position:
\be
g =  \left( \begin{array}{cc} 1 & 0 \\
0 & 0 \end{array} \right).
\label{g}
\ee
The coefficient $\gamma$ is 
determined by the rates of all neutrino reactions in the plasma;
the calculations can be found in 
refs.~\cite{harris92,enqvist92,dolgov97,dolgov99}.
However this expression is not satisfactory for our purpose, in
particular, because it does not conserve particle number in the
case of elastic scattering. To this end we need the exact equation
for the density matrix of oscillating neutrinos derived in
refs.~\cite{dolgov81,sigl93}. However the Fermi blocking factors
will be neglected in what follows. The contribution to the r.h.s.
of kinetic equations from elastic scattering of neutrinos on some
other leptons, $l$, in the plasma is
given by the anticommutators:
\be
\left( {d\rho (p_1) \over dt}\right)_{el} = 
-(A^2_{el}/2) \left(
f_l(p_2) \{g^2, \rho(p_1)\} - f_l(p_4) \{g^2, \rho(p_3)\} \right)
\label{drel}
\ee
where $A_{el}$ is the amplitude of elastic scattering properly 
normalized to give a correct result for the diagonal matrix elements.
It is assumed that the leptons $l$ are not oscillating, otherwise,
if the latter are oscillating neutrinos, the matrix structure of the
result would be much more complicated but in the first approximation
we can use the expression (\ref{drel}).
The integration over momenta of all particles except for 1 are assumed,
namely the following integration of the r.h.s. should be done:
\be
{1\over 2E_1} \int {d^3 p_2 \over (2\pi)^3 2E_2}
{d^3 p_2 \over (2\pi)^3 2E_2} {d^3 p_2 \over (2\pi)^3 2E_2}
(2\pi)^4 \delta^4 \left( p_1+p_2-p_3-p_4\right)
\label{int}
\ee
The amplitude of elastic scattering with proper symmetrization
factors can be taken from tables of ref.~\cite{dolgov97}.

Neutrino annihilation is described similarly:
\be
\left( {d\rho (p_1) \over dt}\right)_{ann} = 
-(A^2_{ann}/4) \left[
\{g, \rho(p_1)g \bar\rho(p_2) \} +\{g, \bar\rho(p_2)g \rho(p_1) \}
- f_l (p_3) f_{\bar l} (p_4) \right]
\label{drann}
\ee

It is convenient to introduce real and imaginary parts of
the non-diagonal components of neutrino density matrix:
\be
\ras = \rsa^* = R +i\,I
\label{rhoas}
\ee
where $a$ and $s$ mean respectively ``active'' and ``sterile''. 

Now the kinetic equations describing evolution of density matrix of
oscillating neutrinos can be written as:
\be
\dot \raa (p_1) &=& -F I - 
A^2_{el} \left[ \raa (p_1) f_l(p_2) - \raa (p_3) f_l(p_4)\right]
\nonumber \\
&& -A^2_{ann} \left[\raa(p_1) \bar\raa (p_2)-f_l(p_3)f_{\bar l}(p_4)
\right],
\label{draa} \\
\dot \rss (p_1) &=& F I,
\label{drss} \\
\dot R (p_1) &=& WI - (A^2_{el}/2) \left[ R(p_1) f_l(p_2)-
R(p_3) f_l(p_4) \right]
\nonumber \\
&& -(A^2_{ann}/4) \left[\raa (p_1) \bar R (p_2) +\bar\raa (p_2) R(p_1)
\right],
\label{dR} \\
\dot I (p_1) &=& -WR - (F/2) \left(\rss -\raa \right)
-(A^2_{el}/2) \left[ I(p_1) f_l(p_2) \right.
\nonumber \\
&& \left. -I(p_3) f_l(p_4) \right]- 
(A^2_{ann}/4) \left[\raa (p_1) \bar I (p_2) +\bar\raa (p_2) I(p_1)
\right].
\label{dI} 
\ee
Here we use notations of ref.~\cite{dolgov99}, so that 
\be
F = \dm \sin 2\theta /2E 
\label{F}
\ee
and 
\be
W = \dm \cos 2\theta /2E + C_l (G_F^2 T^4 E /\alpha)
\label{W}
\ee
where $\alpha = 1/137$, $G_F = 1.166\cdot 10^{-5} {\rm GeV}^{-2}$,
and the constant $C_l$ depends upon the neutrino flavor,
$C_e = 0.61$ and $C_{\mu,\tau} = 0.17$~\cite{notzold88}. We neglected 
the term related to charge asymmetry of the plasma, it may be a good
approximation in non-resonant case. 

In the cosmological FRW background the time derivative in the 
l.h.s. of kinetic equations goes into 
$d/dt \rar \partial /\partial t - Hp\partial /\partial p$, where
$H$ is the Hubble parameter expressed through the 
thermal energy density as
\be
{3 H^2 m_{Pl}^2 \over 8 \pi } = {\pi ^2 g_* \over 30}\, T^4  
\label{h2}
\ee
The factor $g_* = 10.75$ is the number of relativistic species in 
the cosmic plasma. 

We will introduce the new variables:
\be
x=m_0 /T\,\, {\rm and}\,\,   y=p/T\,,
\label{xy}
\ee
where the dimensional normalization  factor $m_0$ is taken 
equal to 1 MeV.
It is a good approximation to assume that the temperature evolves
as the inverse cosmic scale factor, $T\sim 1/a(t)$. In other words,
$\dot T =-HT$ and thus the differential operator 
$(\partial_t -Hp\partial_p)$ transforms into $Hx\partial_x$.

Since the oscillation rate is much faster than the reaction rate 
(in other words, the terms $\sim F,W$ are typically larger than 
the terms related to reactions) the equation~(\ref{dI}) can be 
solved as
\be
R = {F\over 2W} \, \raa
\label{R}
\ee 
This permits to express the imaginary part $I$ through $\raa$ using
eq.~(\ref{dR}) and to substitute the result into eq.~(\ref{draa}).
If we assume that $\rss$ is small in comparison with $\raa$ so 
it may be neglected, the eq.~(\ref{draa}) becomes a closed
equation for a single unknown function $\raa$. If elastic scattering
of active neutrinos is sufficiently strong (its rate is 
approximately an order of magnitude larger than the rate of 
annihilation), then one may assume that the distribution of active
neutrinos is close to kinetic equilibrium with an effective chemical
potential. In other words, the anzats $\raa = C(x) \exp (-y)$
is a good approximation. In this case both sides 
of eq.~(\ref{draa}) can be integrated over $d^3y$ so that the 
contribution of elastic scattering disappears and the following 
ordinary differential equation describing evolution of $C(x)$
is obtained:
\be
{dC\over dx} =-{K_l\over x^4} \left[ C^2 -1 + 
{C^2 \over 288} \left( 6I_2 +I_1^2 \right)
+ {C \over 96}\,{6 +16(g_L^2 +g_R^2)/3 \over 1+2g_L^2 +2g_R^2}\,
\left( 6I_2 -I_1^2 \right) \right]
\label{dcdx}
\ee
where the constant $K_l$ is given by
\be
K_l={8G_F^2 \left( 1+2g_L^2 +2g_R^2\right) \over \pi^3 }
\label{K}
\ee
and
\be
I_n = \int_0^\infty dy\,y^3 e^{-y}\left({F\over W}\right)^n
=\tan 2\theta
\int_0^\infty {dy\,y^3 e^{-y}\over 
\left( 1+ \beta_l\, y^2 x^{-6} \right)^n },
\label{In}
\ee
with
\be
\beta_e = {2.34\cdot 10^{-8} \over \dm \cos 2\theta}\,\,\,
{\rm and}\,\,\, 
\beta_{\mu,\tau} = {0.65\cdot 10^{-8} \over \dm \cos 2\theta},
\label{beta}
\ee
see eqs.~(\ref{F},\ref{W}).

The coupling constants of neutrinos to electrons are 
\be
g_L^2 = (0.5 \pm \sin^2 \theta_W)^2\,\,\, {\rm and}\,\,\,
g_R^2 = \sin^4 \theta_W
\label{glgr}
\ee
The signs ``$\pm$'' refer to $\nue$ and $\nu_{\mu,\tau}$ 
respectively. With $\sin^2 \theta_W = 0.23$ we obtain
$g_L^2 +g_R^2 = 0.5858$ for $\nue$ and $g_L^2 +g_R^2 = 0.1258$
for $\nu_{\mu,\tau}$. Correspondingly $K_e = 0.17$ and
$K_{\nu,\tau} = 0.098$. We assumed that $F/W \ll 1$ and thus the
term $\sim (F/W)^2 dC/dx$ was neglected. It is a good 
approximation even for not very weak mixing.

Eq. (\ref{dcdx}) can be solved analytically if 
$|\delta|=|1-C| \ll 1$:
\be
\delta = \int_0^x dx_1 D_l(x_1) \exp \left[-{2K\over 3}\left(
{1\over x_1^3} - {1\over x^3} \right)\right]
\label{delta}
\ee
where
\be
D(x)_l = K\,x^{-4}\left[ a \left( 6I_2 + I_1^2 \right)
+ b_l \left( 6I_2 -I_1^2 \right)\right]
\label{Dx}
\ee
with $a = 3.47\cdot 10^{-3}$, $b_e = 4.37\cdot 10^{-2}$,
and $b_{\mu,\tau} = 5.55\cdot 10^{-2}$. This expression determines
the freezing temperature of annihilation of active neutrinos, 
$T^{(f)}_l = (3/2K_l)^{1/3}$ MeV.

The increase of the total number density of oscillating neutrinos
\be
\Delta n \equiv \int {d^3\, y \over (2\pi)^3}\Delta (\raa + \rss)
\label{dntot}
\ee 
can be found from the sum
of equations (\ref{draa}) and (\ref{drss}) and is given by
\be
\left({\Delta n \over n_{eq}}\right)_l = 
\int_0^\infty dx\, D_l(x)
\label{deltan}
\ee
All integrals can be taken analytically, first over $dx$ and then
over $dy$. Finally we obtain:
\be
\left({\Delta n \over n_{eq}}\right)_l =
{\pi K_l \over \sqrt {\beta_l}} 
\left({ \sin 2\theta \over \cos 2\theta}\right)^2 
\left[ a_l + b_l + (5.14/6) (a_l - b_l) \right]
\label{dnfin}
\ee
This number should be smaller than the upper limit for extra 
neutrino species, $\Delta N_\nu$, permitted by BBN.
In particular, for electronic neutrinos we obtain (for small
mixing angles):
\be
\dm\,\sin^4 2\theta|_{\nue} < 5\cdot 10^{-4} \Delta N_\nu^2,
\label{dne}
\ee
while for $\nu_{\mu,\tau}$ the result is surprisingly stronger:
\be
\dm\,\sin^4 2\theta|_{\nu_\mu,\nu_\tau} < 
3.3\cdot 10^{-4} \Delta N_\nu^2
\label{dnmu}
\ee
The result makes sense only if $\Delta N <1$. Otherwise, even
if the excitation of sterile neutrinos is complete, i.e. they
reached equilibrium abundance, their maximum contribution to
the effective number of neutrinos would be unity.
The bounds (\ref{dne},\ref{dnmu}) are approximately an order of
magnitude weaker than those obtained in ref.~\cite{enqvist92}
and an order of magnitude stronger than found in 
ref.~\cite{barbieri90}.
In the previous papers the bounds on oscillation parameters of 
$\nue$ was stronger than those for $\num$ and $\nut$
and it was related to a faster production
of $\nue$ by charged currents. However, as we saw above, there
is a competing contribution related to the refraction index of
active neutrinos: it is larger for $\nue$, see 
eqs.~(\ref{W},\ref{beta}).
Hence the oscillations of $\nu_{\mu,\tau}$ are less suppressed 
at high temperatures
and their sterile partners are more efficiently produced. 

We neglected here a distortion of neutrino spectrum by the
oscillations. In contrast to the limits (\ref{dne},\ref{dnmu})
that are effective even for very small mixings, because the
smallness of mixing can be compensated by the efficient production 
of $\nu_s$ at the early stages, the distortion of spectrum of $\nue$
could only be developed sufficiently late, at $T< 2$ MeV and 
the effect could be essential for sufficiently large $\sin \theta$.
One possible form of spectrum distortion is a generation of
an effective chemical potential of the same sign for $\nu$ and 
$\bar\nu$, while kinetic equilibrium shape of the distribution 
survives. This effect dominates for sufficiently large $\dm$.
It was estimated in ref.~\cite{barbieri90}. 
A deviation of electronic neutrinos from kinetic equilibrium
is essential for smaller mass difference. It was accurately
calculated for $\dm \leq 10^{-7} {\rm eV}^2$ 
in the papers~\cite{kirilova97}. Spectral distortion may have 
a very strong effect on primordial abundances, so the limits
on the oscillation parameters could be sensible even if the
data permits $\Delta N_\nu >1$.

Electronic neutrinos are in a good thermal contact with the rest
of the cosmic plasma till 
$T=2$ MeV{\footnote{The reaction rate is proportional to neutrino
momentum $p$, see e.g. ref.~\cite{dolgov99} where this issue is 
discussed. Thus the equilibrium is maintained down to a smaller 
$T$ for larger $p$ and vice versa.}}.
Above this temperature one 
may assume that the spectrum of $\nue$ is close to the equilibrium
Fermi-Dirac form (in fact we will take Boltzmann approximation,
$f_\nu = f^{(eq)}\exp (-E/T)$). Below 2 MeV $\nue$ may be considered 
as free and the impact of medium enters only through the refraction
index. One can check that for $\dm > 5\cdot 10^{-6}$ eV$^2$ the
vacuum term in effective potential (\ref{W}) dominates and 
neutrino oscillations are close to the vacuum ones. 
The number density of electronic neutrinos in this case is
equal to:
\be
\rho_{\nue,\nue} = \left( c^4 + s^4  +2 c^2 s^2 
\cos \dm t /2E \right)\, f^{(eq)} (E),
\label{rhonue}
\ee
where $s=\sin 2\theta$ and $c=\cos 2\theta$. The last term in 
this expression quickly oscillates at nucleosynthesis time scale,
so it can be neglected. This term would give a non-negligible
contribution to neutrino spectrum for $\dm < 10^{-8}$ eV$^2$, but
in this case the matter effects should be taken into account.
From eq.~(\ref{rhonue}) follows that the effective chemical
potential of $\nue$ is 
\be
\xi = \mu/T = \ln \left( 1 - s^2 /2 \right)
\label{xi}
\ee
As is argued in ref.~\cite{barbieri90} such a distortion of
equilibrium neutrino spectrum is equivalent to a renormalization 
of the Fermi coupling constant $G_F \rar G_F [1+\exp (\xi)]/2$.
Since $\xi <0$ the freezing of $n/p$-ratio would take place at 
higher $T$ and more $^4 He$ would be produced. This confines 
the mixing of $\nue$ with $\nu_s$ to the region:
\be
\sin^2 2 \theta < 0.32 \,\Delta N_\nu
\label{s22}
\ee
for all $\dm > 5\cdot 10^{-6}$ eV$^2$. 

For smaller mass differences matter effects cannot be neglected
and the effective mixing angle would depend on neutrino energy:
\be
\sin 2\theta_{eff} = { \sin 2\theta \over 1 + 1.22\,E^2 T^4
(G_F\,^2 /\alpha) |\dm|^{-1 } }
\label{thetaeff}
\ee
This would modify the energy spectrum of $\nue$ and change the
frozen value of neutron-to-proton ratio. Detailed calculations
for $|\dm|< 10^{-7}$ eV$^2$ can be found in the 
papers~\cite{kirilova97}. 

The observational limit on $ \Delta N_\nu$ was analyzed in a 
recent review~\cite{tytler00} with the conclusion 
that $\Delta N_\nu<0.2$. Thus a large mixing of $\num$ with a
sterile partner proposed for explanation of the atmospheric
$\num$-deficit~\cite{sk} is excluded in non-resonant case. 
However, as is argued in ref.~\cite{lisi99}, one should be 
cautious in the data analysis and probably the safer limit
is  $\Delta N_\nu<1$. If this is true an arbitrary strong mixing 
of active neutrinos $\num$ and $\nut$ with one sterile companion 
would be permitted by BBN. In the case of $\nue$-oscillations 
the distortion of $\nue$ spectrum could be essential and in this 
case an interesting bound can be observed even if $\Delta N_\nu =1$
is permitted. 
However if one admits existence of one sterile neutrino,
it is natural to assume that there are three of them, as e.g. 
in the case of Dirac-Majorana mass mixing~\cite{dolgov81}.
In this case the mixing angles should be rather strongly bound 
by BBN, but detailed analysis with correct description of 
decoherence effects is necessary.

\bigskip
{\bf Acknowledgement.} I am grateful to M. Chizhov, S. Hansen,
and D. Kirilova  for discussions and comments.

\end{document}